\documentclass[conference]{IEEEtran}
\usepackage{amssymb}
\usepackage{amsmath,amsfonts}
\usepackage{algorithmicx}
\usepackage{amsthm}
\usepackage{algorithm}
\usepackage{array}
\usepackage{textcomp}
\usepackage{stfloats}
\usepackage{url}
\usepackage{verbatim}
\usepackage{graphicx}
\usepackage{cite}
\usepackage{algpseudocode}
\usepackage{bm}
\usepackage{graphicx}
\usepackage{subfigure}
\newcommand{\RNum}[1]{\uppercase\expandafter{\romannumeral #1\relax}}

\theoremstyle{definition}

\newtheorem{lemma}{Lemma}

\theoremstyle{remark}

\def\BibTeX{{\rm B\kern-.05em{\sc i\kern-.025em b}\kern-.08em
    T\kern-.1667em\lower.7ex\hbox{E}\kern-.125emX}}

\begin{document}

\title{ Performance Analysis of Hybrid Cellular and Cell-free MIMO Network 
}

\author{{Zhuoyin Dai{*},  Jingran Xu{*}, Xiaoli Xu{*}, Ruoguang Li{*} and
Yong Zeng{*$\dagger$}
  }  \\
  {{*}National Mobile Communications Research Laboratory, Southeast University, Nanjing 210096, China} \\
  {{†}Purple Mountain Laboratories, Nanjing 211111, China} \\
  {Email: \{zhuoyin\_dai, jingran\_xu, xiaolixu, ruoguangli, yong\_zeng\}@seu.edu.cn}




 
} 

 \maketitle

\begin{abstract}
Cell-free wireless communication is envisioned as one of the most promising 
network architectures, 
which can achieve stable and uniform 
communication performance while improving the system energy and spectrum efficiency.
The deployment of cell-free networks is envisioned to be a long-term 
evolutionary process, in which cell-free access points (APs) will be gradually introduced 
into the communication network and collaborate with the existing cellular 
base stations (BSs).  
To further explore the performance limits of hybrid cellular and cell-free networks, 
this paper develops a   hybrid network model based on stochastic geometric toolkits,
which reveals the coupling of the signal and interference from both the cellular 
and cell-free networks. Specifically, 
the conjugate beamforming is applied in hybrid cellular and cell-free networks, which enables user equipment (UE) 
to benefit from both cellular BSs and cell-free APs.
The aggregate signal received from the hybrid network is approximated via
moment matching, and  coverage probability  
is characterized by deriving the Laplace transform of the interference. 
The analysis of signal strength and coverage probability is verified by extensive simulations. 

\end{abstract}

   

\section{Introduction}
The future wireless communication networks will witness a proliferation of mobile 
applications and unprecedented growth in wireless data. 
The realization of higher spectrum and energy efficiency with 
superior   costs remains a challenging 
issue in current research. As one of the most prominent wireless technologies 
proposed in recent years, cell-free   network
is considered as a promising network architecture
in the beyond fifth-generation (B5G) and sixth generation (6G) mobile communication system 
\cite{ngo2017cell }.
Different from traditional cellular systems, cell-free network is a user-centric 
coverage architecture that discards the traditional concept of cellular boundaries \cite{wangdongming2020performa}. 
The  central processing unit (CPU) controls the access points (APs) to
cooperate  to provide services to 
user equipment (UE) on the same time-frequency resources,
thus realizing higher spatial multiplexing \cite{nayebi2017precoding,zhuoyindai2023characterizing}. 
Cell-free network improves the energy and spectral efficiency of the system, 
and effectively reduces the 
performance gaps between UE by ensuring that there are spatially 
short-range APs that provide stable services to UE.

However, the deployment of cell-free systems in existing commercial mobile 
networks still faces serious challenges. First, the construction of 
cell-free systems requires the deployment of the distributed APs throughout 
the network and the construction of the corresponding fronthaul links, 
which brings heavy time costs and deployment expenses. Second, 
simply introducing the cell-free system without cooperation will inevitably cause mutual interference 
with existing cellular systems, significantly limiting the system performance. 
Therefore, the deployment of cell-free systems is bound to be a long-term 
evolutionary process, and hybrid cellular and cell-free cooperation networks are both 
a necessity and a desirable choice for 
B5G and 6G.

Some recent works have investigated the performance analysis and resource allocation  
of hybrid cellular and cell-free 
networks. A cell-free and legacy cellular coexistence system deployed on the 
existing system architecture, as well as the corresponding precoding, 
power control, etc., are outlined in \cite{kim2022howwill}. 
With appropriate UE association criteria and coordinated beamforming, hybrid cell-free and small cell systems 
can provide superior downlink rates for static and dynamic UE than the single 
architecture \cite{Elhoushy2021towards}. However, the above works do not take into 
account the impact of the spatial distribution of  base stations (BSs), APs, and UE  on 
network performance. 

Due to the densification and irregularity of wireless 
node distributions in the network, traditional grid-based deployment models 
are difficult to reflect the practical system performance. 
Stochastic geometry models the spatial distribution of wireless nodes 
with point processes and can effectively characterize the lower bound of 
the actual system performance. There have been some works 
using stochastic geometry 
to analyze the performance of cell-free networks
and coordinated multiple points (CoMP) communication  in terms of 
energy efficiency (EE) \cite{Papazafeiropoulos2020performance},
power control \cite{hoang2018cellfree} 
and channel hardening analysis \cite{chen2018channelhardening}, etc.
However, there is still a lack of work related to the   characterization 
of hybrid cellular and cell-free networks. 
In addition, the existing stochastic geometry-based heterogeneous network studies, 
which separate different network layers from each other 
\cite{chun2015modeling}
, are not applicable 
to the analysis of hybrid cellular and cell-free networks.

To gain some insights of the performance limit of the hybrid network, 
this paper develops a stochastic geometry-based model 
for hybrid cellular and cell-free networks, 
which reveals the coupling of the signal and interference from both 
the cellular and cell-free networks. 
However, the aggregate signals from the BSs and APs with conjugate beamforming 
make it difficult to characterize the distribution of the signal strength 
and the corresponding signal-to-interference plus
noise ratio (SINR). 
To tackle this issue, we first derive the closed-form expressions for the average   
signal strength and interference power from the APs, 
and then the aggregate signal strength distribution 
is approximated via moment matching. Finally, the  coverage probability
is characterized on the basis of the Laplace transform of the system interference power.
The analysis of network coverage probability 
is verified by extensive simulations, and it can be used to guide the network 
deployment and interference management in the hybrid cellular and cell-free networks.

\section{System Model}
\begin{figure}[!t]
 \centering
   {\includegraphics[width=0.8\columnwidth]{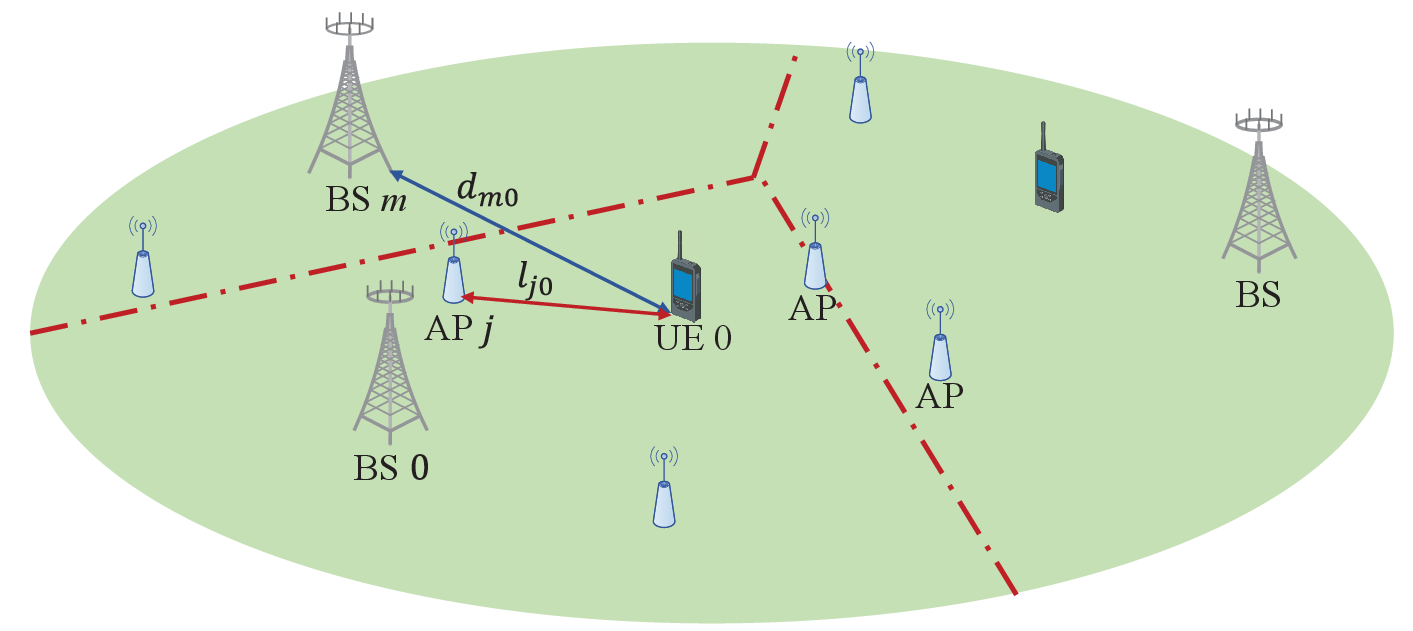}}%
 \caption{Hybrid cellular and cell-free network.}
 \label{fig:CCCN}
\end{figure}

As shown in Fig. \ref{fig:CCCN},  a hybrid cellular and cell-free network is considered
in this paper.  
The locations of BSs, cell-free APs, and
single-antenna  UE  
are modeled by independent homogeneous Poisson point processes (HPPP) $\Lambda_{B}$, 
$\Lambda_{A}$ and $\Lambda_{U}$, 
with   density $\lambda_{B}$/$\mathrm{km}^2$, $\lambda_{A}$/$\mathrm{km}^2$ and $\lambda_{U}$/$\mathrm{km}^2$, respectively.
Each BS is equipped with $N_B$ antennas, while each AP  is equipped with $N_A$ antennas.
Considering the different configurations  that AP and BS can support, $P_A$ and $P_B$ 
denote the maximum downlink power of AP and BS with  $P_A < P_B$.  
In the network, $d_{mi}$ and $l_{ji}$ denote the distance between BS $m$ and UE $i$,
and that between AP $j$ and UE $i$, respectively. 
We consider a typical UE, referred as UE 0, which is jointly served by the closest BS, 
named BS 0, with the distance $d_{00}$, and all the cell-free APs in $\mathcal{A}$.
The channel vector between BS $m$ and 
UE $i$  is denoted by $\mathbf{h}_{mi} \triangleq [h_{mi,1},...,h_{mi,N_B} ] \in 
\mathbb{C}^{N_B\times 1 }$, while the channel vector between AP $j$ and UE $i$ is 
$\mathbf{g}_{ji} \triangleq [g_{ji,1},...,g_{ji,N_A} ]  \in \mathbb{C}^{N_A\times 1 }$. 
The channel model consisting of distance-dependent large-scale fading and 
random small-scale fading is considered as 
\begin{equation}\label{eq:hmi}
  \mathbf{h}_{mi}=\beta_{mi}^{\frac{1}{2}}\bm{\zeta}_{mi},m \in \omega_B,
\end{equation} 
\begin{equation}\label{eq:gji}
  \mathbf{g}_{ji}=\delta_{ji}^{\frac{1}{2}}\bm{\xi}_{ji},j \in \omega_A,
\end{equation}
where $\beta_{mi}$ and $\delta_{ji}$  are path loss of the channel with  
$\beta_{mi}=\beta_0 d_{mi}^{-\alpha_1}$ and $\delta_{ji}=\delta_0 l_{ji}^{-\alpha_2}$.
$\omega_B$ and $\omega_A$ are the sets of all the BSs and APs, respectively. 
The small-scale fading   in both  $\bm{\zeta}_{mi}$ and $\bm{\xi}_{ji}$ are 
independent and identically distributed (i.i.d.) $\mathcal{CN}(0,1)$  random variables (r.v.s).

The entire network area is represented by $\mathcal{A}$.

During the downlink transmission,  UE in the same cell will be served 
by the same BS.
In addition, both BSs and APs select conjugate beamforming  
in order to obtain low computational complexity and good performance,
and also to avoid channel state information (CSI) interactions between APs \cite{ngo2017cell}.
Therefore, the signal transmitted by BS $m$  is 
\begin{equation}
  \mathbf{x}_{B,m}=\sqrt{P_B \eta_B} \sum_{n \in \phi_{B,m}}
  \frac{\mathbf{h}_{mn}}{\|\mathbf{h}_{mn} \|}q_{n},
\end{equation}
where $\phi_{B,m}$ denotes the set of UE 
that are served by BS $m$, 
and $q_{n} \sim \mathcal{CN}(0,1)$ denotes the  information-bearing symbols for 
UE $n$.  $\eta_B$ denotes the power constraint parameter with
$\mathbb{E}[  \mathbf{x}_{B,m}^H \mathbf{x}_{B,m} ]=P_B$.
For convenience, $\eta_B$ is expressed as the average number of users per BS, i.e.,
$\eta_B=\frac{1}{|\bar{\phi}_{B}|}=\frac{\lambda_B}{\lambda_U}$,
where $|\bar{\phi}_B|$ denotes the average of $|\phi_{B,m}|$ for any $m$.

Denote the set of UE in the network as $\phi_U$. Therefore, 
the corresponding downlink signal from each cell-free AP $j$   is 
\begin{equation}
  \mathbf{x}_{A,j}=\sqrt{P_A \eta_A} \sum_{i\in \phi_{U}} \frac{ \mathbf{g}_{ji}}{\|\mathbf{g}_{ji}\|}q_i,
\end{equation}
where $\eta_A$ denotes the power constraint parameter.
Note that the averaged number of  UE in $\phi_{U}$ is  
$\mathbb{E}[| \phi_{U}|]=\bar{U}=\lambda_U|\mathcal{A}|$, $\eta_A$ can be 
denoted as $\eta_A=\frac{1}{\bar{U}}$
to ensure $\mathbb{E}[  \mathbf{x}_{A,j}^H \mathbf{x}_{A,j} ]=P_A$.

For any UE $i$ in the network, $i^{*}$ is denoted as the index of the associated and
nearest BS providing the service.
With the collaboration of cell-free APs and  cellular BSs, 
the downlink signal received by the typical UE 0 is 
\begin{equation}\label{eq:y0}
  \resizebox{1\hsize}{!}
  {$\begin{aligned}
   y_0&=\sum_{m\in \omega_B}\mathbf{h}_{m0}^H \mathbf{x}_{B,m} + 
  \sum_{j\in \omega_{A}}\mathbf{g}_{j0}^H\mathbf{x}_{A,j}+n_0
\\ &=\underbrace{\sqrt{P_B \eta_B}\|\mathbf{h}_{00} \|}_{S_{0B}}q_0
+\underbrace{\sum_{j\in \omega_{A}} \sqrt{P_A \eta_A}   \|\mathbf{g}_{j0}\|}_{S_{0A}}q_0+
\\ &  \underbrace{\sum_{i\in \phi_{U}\backslash_{0}}\! \Big(
  \sqrt{P_B\eta_B}\mathbf{h}_{i^{*}0}^H \frac{\mathbf{h}_{i^{*}i}}
  {\|\mathbf{h}_{i^{*}i} \|}
  \!+\!\sqrt{P_A \eta_A}  \sum_{j\in \omega_{A} }
  \mathbf{g}_{j0}^H \frac{ \mathbf{g}_{ji}}{\|\mathbf{g}_{ji}\|} \Big) }_{I_{U}}q_{i}+n_0,
  \end{aligned} $}
\end{equation}
where the first term $S_{0A}$ and the second term $S_{0B} $   represent the desired signals from 
BS 0 and    cell-free APs, respectively. 
The total interference   is shown in the third term $I_{U}$.
Each term in $I_{U}$ consists of the signal sent by cellular  BSs   
and cell-free APs to
any other UE. The last term $n_0 $ denotes  the additive white Gaussian noise (AWGN)
with power $\sigma^2$. 

Based on  (\ref{eq:y0}), the  interference power caused   by the signal intended to 
UE $i$ is given by
\begin{equation}\label{eq:Ii}
  I_{i}=\Big|
    \sqrt{P_B\eta_B}\mathbf{h}_{i^{*}0}^H \frac{\mathbf{h}_{i^{*}i}}
    {\|\mathbf{h}_{i^{*}i} \|}
    \!+\!\sqrt{P_A \eta_A}  \sum_{j\in \omega_{A} }
    \mathbf{g}_{j0}^H \frac{ \mathbf{g}_{ji}}{\|\mathbf{g}_{ji}\|} \Big|^2.
\end{equation}

In interference $I_{i}$, BS channel vector $\mathbf{h}_{i^{*}0}$ is independent of 
channel vector $\mathbf{g}_{j0}, \forall j \in \omega_{A} $. Meanwhile, the channel gains 
$\mathbf{g}_{i0}$ and $\mathbf{g}_{j0}$
from different APs are also independent of each other, $\forall i \neq j $.
Taking into account the law of large number   and the mutual independence of 
channels as well as beamforming vectors between different BSs and APs, $I_{i}$ can be approximated as
\begin{equation}\label{eq:Iiapp}
  I_{i}\approx P_B\eta_B \Big|\mathbf{h}_{i^{*}0}^H \frac{\mathbf{h}_{i^{*}i}}
  {\|\mathbf{h}_{i^{*}i} \|}\Big|^{2}+P_A \eta_A\sum_{j\in \omega_{A}}
  \Big| \mathbf{g}_{j0}^H \frac{ \mathbf{g}_{ji}}{\|\mathbf{g}_{ji}\|}\Big|^{2}.
\end{equation}

The approximation   in (\ref{eq:Iiapp}) 
indicates that $I_{i}$ can be expressed in the form of the sum of the 
power of the signal from the associated BS $i^*$ and  each AP. Further,
by classifying the interference into  intra-cell interference $I_{B0}$, 
inter-cell interference $I_{B}$,
and  interference $I_{A}$ due to the APs, 
the total interference $\sum_{i\in \phi_{U}\backslash\{0\}}I_{i}$ can be rewritten as 
\begin{equation}
  \sum_{i\in \phi_{U}\backslash\{0\}}I_{i}=I_{B0}+I_{B}+I_{A},
\end{equation}
where 
\begin{equation}\label{eq:SI}
  \begin{aligned}
    I_{B0}&= P_B\eta_B \sum_{n\in \phi_{B,0}\backslash \{0\}} 
    \Big| \mathbf{h}_{00}^H \frac{\mathbf{h}_{0n}}{\|\mathbf{h}_{0n} \|}\Big|^2,\\
    I_{B}&= P_B \eta_B\sum_{m\in \omega_B \backslash \{0\}} 
    \sum_{n\in \phi_{B,m}}
    \Big| \mathbf{h}_{m0}^H \frac{\mathbf{h}_{mn}}{\|\mathbf{h}_{mn} \|} \Big|^2,\\
    I_{A}&=P_A \eta_A  \sum_{i\in \phi_{U}\backslash\{0\}}\sum_{j\in \omega_{A}} \Big|   
    \mathbf{g}_{j0}^H \frac{ \mathbf{g}_{ji}}{\|\mathbf{g}_{ji}\|}  \Big|^2.
  \end{aligned}
\end{equation}


Therefore,
the corresponding received SINR at UE 0 
is approximately  expressed as 
\begin{equation}\label{eq:SINR}
  \Omega=\frac{S_0}{I_{B0}+I_{B}+I_{A}+\sigma^2},
\end{equation}
where 
\begin{equation}
  \resizebox{1\hsize}{!}
  {$\begin{aligned}
  S_0&=\big( S_{0B}+S_{0A}\big)^2 = \Big( \sqrt{P_B \eta_B}\|\mathbf{h}_{00} \|   
  +\sqrt{P_A \eta_A}  \sum_{j\in \omega_{A}}   \|\mathbf{g}_{j0}\| \Big)^2.
  \end{aligned}$}
\end{equation}


\section{Analysis of Signal Strength and Interference}
In this section, 
the statistical distributions of the received signal and interference power 
are characterized.
The signal strength is approximated via moment matching, 
with its first-   and second-order moments derived in closed-form. 
The intra- and inter-cell interference caused by BSs 
are approximated as a Gamma r.v. and a weighted sum of Gamma  r.v.s, 
respectively. The average interference caused by cell-free APs is derived in closed-form.
In addition, the performance of the network 
can be obtained by analyzing a typical UE 0 according to Slivnyak's theorem \cite{andrews2011tractable}.

\subsection{Analysis of   Channel Distribution}



Based on the channel model of BSs and APs,
the power of the channel to UE $i$ for the $m$th BS and the $j$th AP 
can be respectively given by
\begin{equation}\label{eq:hmi2}
  |\mathbf{h}_{mi}|^2=\beta_{mi}\bm{\zeta}_{mi}^{H}\bm{\zeta}_{mi},
\end{equation}
\begin{equation}\label{eq:gji2}
  |\mathbf{g}_{ji}|^2=\delta_{ji}\bm{\xi}_{ji}^{H}\bm{\xi}_{ji}.
\end{equation}

Since all the entries in both $\bm{\zeta}_{mi}$ and $\bm{\xi}_{ji}$ 
follow the i.i.d. $\mathcal{CN}(0,1)$,
$\bm{\zeta}_{mi}$ and $\bm{\xi}_{ji}$  are isotropic vectors in $N_B$ and $N_A$
dimensions respectively \cite{isotropic}.

Note that for the isotropic vector $\mathbf{x}\in \mathbb{C}^{N\times 1}$ with each entry
following i.i.d. $\mathcal{CN}(1,\delta^2)$, $\mathbf{x}^H\mathbf{x}$ is the sum of i.i.d.
variables $\Gamma(1,\delta^2)$, and thus follows $\Gamma(N,\delta^2)$ \cite{health2011multiuser}. 
Therefore, we have $\bm{\zeta}_{00}^{H}\bm{\zeta}_{00}\sim \Gamma(N_B,1) $ and 
$\bm{\xi}_{j0}^{H}\bm{\xi}_{j0}\sim \Gamma(N_A,1)$.
\begin{lemma}
  \label{lem1}
For  the Gamma distributed r.v. $X \sim \Gamma(a,\theta)$ and any $b>0$,
  $Y=bX\sim \Gamma(a,b\theta)$ \cite{moschopoulos1985distribution}.
\end{lemma}

Based on  Lemma \ref{lem1}, the BS and AP channel power  in 
(\ref{eq:hmi2}) and (\ref{eq:gji2}) are distributed according to
\begin{equation}\label{eq:h2sim}
  |\mathbf{h}_{mi}|^2\sim \Gamma(N_B,\beta_{mi}),
\end{equation}
\begin{equation}
  |\mathbf{g}_{ji}|^2\sim \Gamma(N_A,\delta_{ji}),
\end{equation}

\subsection{ Approximation of the Signal Power Distribution}
According to (\ref{eq:h2sim}),
 the power of the nearest associated BS channel $|\mathbf{h}_{00}|^2$
is the sum of $N_B$ i.i.d. variables
following $\Gamma(1,\beta_{00})$, i.e., $|\mathbf{h}_{00}|^2\sim 
\Gamma(N_B,\beta_{00})$. 
For further analysis of the desired signal $S_0$ in (\ref{eq:SI}), Lemma \ref{lem5} about the square root of 
Gamma variable is first introduced.
\begin{lemma}
  \label{lem5}
For  any Gamma distributed r.v. $X \sim \Gamma(k,\theta)$,
the square root $Y$ of $X$ follows the Nakagami distribution
as $Y=\sqrt{X}\sim \mathrm{Nakagami}(m,\omega)$ \cite{huang2016nakagami},
where the parameters are $m=k, \omega=m\theta$.
\end{lemma}

Therefore, the distribution of $\|\mathbf{h}_{00} \| $ is obtained
according to  Lemma \ref{lem5} as 
\begin{equation}
  \|\mathbf{h}_{00} \|=\sqrt{|\mathbf{h}_{00}|^2}\sim \mathrm{Nakagami}(N_B, N_B \beta_{00}),
\end{equation}
while the component of AP channel $ \|\mathbf{g}_{j0} \|$ in $S_0$ has
\begin{equation}
  \|\mathbf{g}_{j0} \|=\sqrt{|\mathbf{g}_{j0}|^2}\sim \mathrm{Nakagami}(N_A, N_A \delta_{j0}), \forall j\in \omega_A. 
\end{equation}

The distribution of  the desired signal $S_0$ is composed of the signals from the associated BS 0 together with all APs.
Considering the random distribution of the cell-free  APs 
in the network,  the following Lemma \ref{lemassLA}
is introduced.

\begin{lemma}
  \label{lemassLA}
  With  the law of large number,  
  the desired signal in $S_0$   due to the APs can be approximated 
  by their average $L_A$ when the number of APs is large and $\alpha_2<4$, i.e., 
\begin{equation}
    \begin{aligned}
    \sqrt{P_A \eta_A}  &\sum_{j\in \omega_{A}}  
  \|\mathbf{g}_{j0}\|  \approx \sqrt{P_A \eta_A}   \mathbb{E}[\sum_{j\in \omega_{A}}  
  \|\mathbf{g}_{j0}\|] \\
   &= \underbrace{\frac{4\pi\sqrt{\rho_A} \lambda_A \delta_{0}^{\frac{1}{2}}}{4-\alpha_2}
   \frac{\Gamma(N_A+\frac{1}{2})}{\Gamma(N_A)}  
   \Big(\frac{|\mathcal{A}|}{\pi}\Big)^{1-\frac{\alpha_2}{4}}}_{L_A},
    \end{aligned}
  \end{equation}     
  where $\rho_A=P_A \eta_A$. 
  The detailed derivation is based on the Campbell Theorem \cite{andrews2011tractable},  
  and will be shown in an extended journal version.
\end{lemma}




From Lemma \ref{lemassLA}, the power expression  $S_0$ for the desired signal 
is simplified as the square of the sum of  a Nakagami r.v. and the constant $L_A$
as $S_0\approx ( \sqrt{P_B \eta_B}\|\mathbf{h}_{00} \|   +L_A )^2$.
Therefore, the following Lemma is introduced.

\begin{lemma}
  \label{lemassshifted}
For any Nakagami r.v. $X\sim  \mathrm{Nakagami}(m,\omega)$, the 
probability density function (PDF) of the 
square of the shifted  Nakagami r.v. $Y=(X+A)^2$ for $Y>A^2$ is 
\begin{equation}\label{eq:fYy}
  \resizebox{1\hsize}{!}
  {$
  \begin{aligned}
  f_{Y}(y)
   & =\!\frac{m^m}{\Gamma(m)\omega^m}\!(\!\sqrt{y}\!-\!A)^{2m\!-\!1}
   \mathrm{exp}\!\big(\!-\!\frac{m}{\omega}(\sqrt{y}\!-\!A)^2\big)y^{\frac{1}{2}}.
\end{aligned}$}
\end{equation}
\end{lemma}

\begin{figure}[!t]
  \centering
    {\includegraphics[width=0.9\columnwidth]{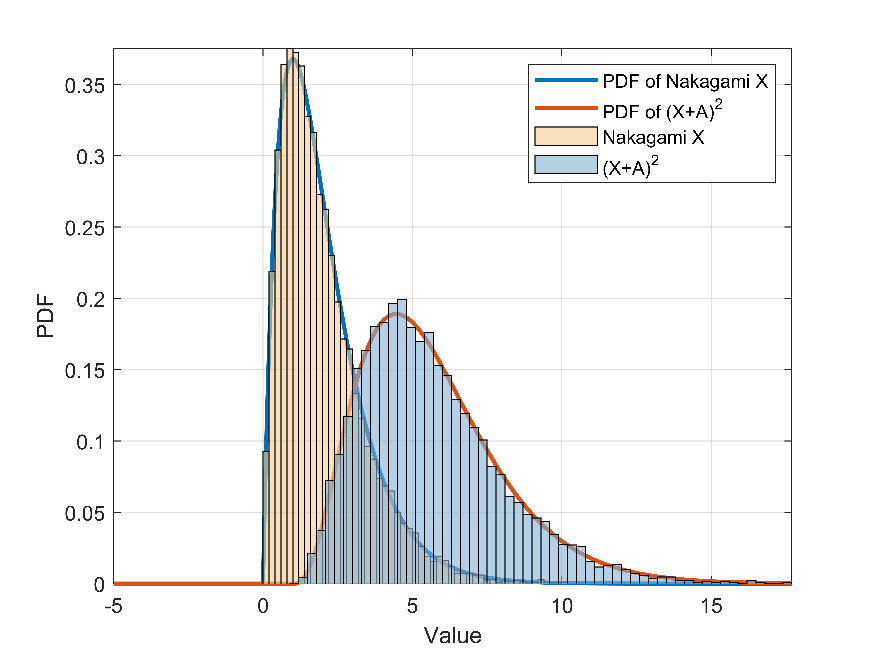}}%
  \caption{The distribution of the 
  square of   shifted  Nakagami r.v. $Y=(X+A)^2$.}
  \label{fig:X+A2}
\end{figure}

The verification of the distribution of $Y=(X+A)^2$ is shown in Fig. \ref{fig:X+A2}.
From (\ref{eq:fYy}) and Fig. \ref{fig:X+A2}, the exact distribution of $S_0$ is difficult to characterize, but the 
corresponding PDF of $S_0$ has a similar structure to that of the Gamma distribution.
Therefore, with given distance of the nearest associated BS 0, 
$S_0$ can be approximated as a Gamma r.v. based on its
first- and second-order moments \cite{Lyu2021IRS}. 
The corresponding Lemma (\ref{lemappS0}) is introduced as follows.
\begin{lemma}
  \label{lemappS0}
  According to the definition of the Gamma r.v. \cite{pishro2014introduction}, the desired signal power
  $S_0$ can be approximated as the   Gamma distribution $\Gamma(k_{S_0},\theta_{S_0})$ with
  \begin{equation}
    \begin{aligned}
      k_{S_0}&=\frac{\big( \mathbb{E}[S_0] \big)^2}{ \mathrm{Var}\{S_0\} }
      =\frac{\big( \mathbb{E}[S_0] \big)^2}{ \mathbb{E}[S_0^2] -\big( \mathbb{E}[S_0] \big)^2},\\
      \theta_{S_0}&=\frac{ \mathrm{Var}\{S_0\} }{\mathbb{E}[S_0] }
      =\frac{ \mathbb{E}[S_0^2] -\big( \mathbb{E}[S_0] \big)^2}{\mathbb{E}[S_0] },
    \end{aligned}
  \end{equation}
where  the first- and second-order moments of $S_0$ are 
\begin{equation}
  \mathbb{E}[S_0]=\rho_B N_B \beta_{00} 
  +2\rho_B^{\frac{1}{2}}L_A\frac{\Gamma(N_B +\frac{1}{2})}{\Gamma(N_B)} \beta_{00}^{\frac{1}{2}} 
  +L_{A}^{2},
\end{equation}
\begin{equation}
  \begin{aligned}
  \mathbb{E}[S_0^2] &=\rho_B^2 N_B(N_B+1)\beta_{00}^2
  +4\rho_B^{\frac{3}{2}}L_A \frac{\Gamma(N_B +\frac{3}{2})}{\Gamma(N_B)}\beta_{00}^{\frac{3}{2}}
  \\ & +6\rho_B L_A^2 N_B \beta_{00}
  +4\rho_B^{\frac{1}{2}}L_A^3\frac{\Gamma(N_B +\frac{1}{2})}{\Gamma(N_B)}\beta_{00}^{\frac{1}{2}}+L_{A}^{4} .
\end{aligned}
\end{equation}

 The detailed derivation is based on the raw moments of Gamma distribution,
 and is omitted here for space.
\end{lemma}

\subsection{Analysis of Interference Power Distribution}

Similar to the case of the desired signal, the power distribution of the interference
is analyzed in this subsection. 
Considering   that conjugate beamforming is applied by both BSs and APs 
in the network,
the Lemma \ref{lem3} for the projection of isotropic channel vectors is introduced. 
\begin{lemma}
 \label{lem3}
Denote $\mathbf{x} \in \mathbb{C}^{N\times1}$ as an isotropic vector with i.i.d.
$\mathcal{CN}(0,\theta)$ entries.
If $\mathbf{x}$ is projected onto an s-dimensional beamforming subspace, the  
power distribution is \cite{muirhead2009aspects}
\begin{equation}
 |\mathbf{x}^H\mathbf{w}|^2\sim \Gamma(s,\theta).
\end{equation} 
\end{lemma}


Based on  Lemma \ref{lem3}, the
power $I_{B0}$ of intra-cell interference in (\ref{eq:SI}) can be approximated as the sum of 
$(|\bar{\phi}_{B}|-1)$ i.i.d. variables following $\Gamma(1,P_B\eta_B\beta_{00})$.
Further, extracting the scale parameter,
the power of intra-cell interference can be rewritten as 
$I_{B0}=P_B\eta_B \beta_{00}\kappa_{B,0}$, where $\kappa_{B,0}\sim 
\Gamma( |\bar{\phi}_{B}|-1,1) $.

For the inter-cell interference, since $\mathbf{h}_{m0}$ is also independent of 
$\mathbf{h}_{mn}$, the interference power of each BS $m\in \omega_B \backslash \{0\}$ 
in $I_{B}$ is approximated as
$\sum_{n\in \phi_{B,m}}
\Big| \mathbf{h}_{m0}^H \frac{\mathbf{h}_{mn}}{\|\mathbf{h}_{mn} \|} \Big|^2
\sim \Gamma( |\bar{\phi}_{B}|,\beta_{m0})$.
Therefore, as the sum of interference from BSs in $\omega_B \backslash \{0\}$,
the inter-cell interference $I_{B}$ can be further expressed as the sum of Gamma variables 
with the same shape parameters  and scale parameters, i.e.,
\begin{equation}\label{eq:IBkappa}
  I_{B}=P_B\eta_B \sum_{m\in\omega_B \backslash \{0\}} \beta_{m0}\kappa_{B,m0},
\end{equation}
where $\kappa_{B,m0}\in \Gamma( |\bar{\phi}_{B}|,1), \forall m \in\omega_B \backslash \{0\}$.

Next, we need to analyze the interference $I_A$ from the APs. 
According to (\ref{lemassLA}) and  (\ref{lem3}), 
the following Lemma about $I_A$ is introduced.
\begin{lemma}\label{lembarIA}
  By   the law of large number,  
  the interference $I_A$   due to the APs is approximated 
  by its average $\bar{I}_{A}$ when the number of APs is large and $\alpha_2<2$, i.e., 
\begin{equation}
  \begin{aligned}
  I_{A}
  &\approx P_A \eta_A \mathbb{E} \bigg[\sum_{i\in \phi_{U}\backslash\{0\}}\sum_{j\in \omega_{A}} \Big|   
    \mathbf{g}_{j0}^H \frac{ \mathbf{g}_{ji}}{\|\mathbf{g}_{ji}\|}  \Big|^2 \bigg]\\
    &= \underbrace{\frac{2\pi \rho_A  \lambda_A \delta_{0}(\lambda_U|\mathcal{A}|-1)}
    {2-\alpha_2}
    \Big(\frac{|\mathcal{A}|}{\pi}\Big)^{1-\frac{\alpha_2}{2}}}_{\bar{I}_{A}}.
  \end{aligned}
  \end{equation} 
  The detailed derivation is based on the Campbell Theorem \cite{andrews2011tractable},  
  and will be shown in an extended journal version.
\end{lemma}

\section{Coverage Probability of Hybrid Cellular and Cell-free Network}
In this section, the  coverage probability  of hybrid cellular and cell-free network 
is analyzed based on the distribution of signal strength and various interference 
components, derived in the preceding section. 
In general, the coverage probability is the complementary cumulative distribution function
(CCDF) of SINR over the overall network, which can be defined as 
\begin{equation}\label{eq:covdef}
  p_{\mathrm{c}}\triangleq\mathbb{P}[ \Omega=\frac{S_0}{I_{B0}+I_{B}+\bar{I}_{A}+\sigma^2} > T],
\end{equation}
where $T$ denotes the target threshold of the SINR $\Omega$.

\subsection{Analysis of   Coverage Probability} 
Taking the distance $d_{00}$ between the typical UE 0 and its associated and nearest 
BS 0 as an r.v., the average  coverage probability in  the network is 
\begin{equation}\label{eq:cov1}
  \begin{aligned}
    p_{\mathrm{c}}&=\mathbb{E}\big[ p_{\mathrm{c}}(d_{00})  \big] = 
    \int_{0}^{\sqrt{\frac{|\mathcal{A}|}{\pi}}} p_{\mathrm{c}}(r) f_{d_{00}}(r) 
    \mathrm{d}r,
  \end{aligned}
\end{equation}
where $f_{d_{00}}(r)$ denotes the PDF of 
the  nearest point distance in PPP. With the
 cumulative distribution function (CDF) of r.v. $d_{00}$ as 
$F_{d_{00}}(r)=1-e^{-  \lambda_B \pi r^2}$ \cite{andrews2011tractable}, there is
\begin{equation}
  \begin{aligned}
    f_{d_{00}}(r)&=\frac{\mathrm{d}F_{d_{00}}(r)}{\mathrm{d}r}& =2\lambda_B \pi r e^{-\lambda_B \pi r^2}.
  \end{aligned}
\end{equation}

First, 
the relationship between intra-cell interference $I_{B0}$ and inter-cell interference
$I_B$ is  analyzed.
Clearly, both $I_{B0}$ and $I_B$ are dependent on the distance $d_{00}$, i.e., 
$I_{B0}$ comes from the associated BS 0 with a distance of $d_{00}$, 
and $I_B$ comes from the other BSs with distances  $d\geq d_{00}$.
However, there is no interaction between  $I_{B0}$ and $I_B$.
Specifically, with the given $d_{00}$, $I_{B0}$ depends  on the distribution of UE
in the cell of BS 0, while $I_B$ depends mainly on the distribution of other BSs 
with a distance no smaller than $d_{00}$.
Therefore, $I_{B0}$ and $I_B$ are independent of 
each other with a given $d_{00}$.
Therefore, the coverage probability can be further expressed as in Lemma \ref{lemcoverage}.
\begin{lemma}
  \label{lemcoverage}
  Considering that the desired signal $S_0$ following the Gamma distribution, i.e.,  $S_0\sim\Gamma(k_{S_0},\theta_{S_0})$
  the network coverage probability in (\ref{eq:covdef}) can be expressed as
\begin{equation}\label{eq:outtolap}
  \begin{aligned}
  &p_{\mathrm{c}}(d_{00})=\mathbb{P}[ S_{0} > T(I_{B0}+I_{B}+\bar{I}_{A}+\sigma^2)]\\ 
  &=\sum_{i=0}^{k_{S_0}-1} \frac{(-1)^i}{i!} \frac{\partial^i}{\partial^i s}
  \Big\{ e^{-s\frac{TI_{e}}{\theta_{S_0}}}
      \mathcal{L}_{Y_{I_{B0}}}(s) 
      \mathcal{L}_{Y_{I_{B}}}(s)   \Big\}_{s=1},
  \end{aligned}
\end{equation}
where the sum of interference from APs and noise is denoted as $I_{e}=\bar{I}_{A}+\sigma^2$. 
The shape parameter $k_{S_{0}}$ is integer. The Laplace transforms 
of   
$Y_{I_{B0}}=\frac{TI_{B0}}{\theta_{S_0}}$ and $Y_{I_{B}}=\frac{TI_{B}}{\theta_{S_0}}$
are 
\begin{equation}\label{eq:YIlap}
  \resizebox{1\hsize}{!}
  {$\begin{aligned}
  &\mathcal{L}_{Y_{I_{B0}} }(s)=\Big(
    1+s\frac{T\rho_B\beta_0 d_{00}^{-\alpha_1}}{\theta_{S_0}}\Big)^{1-|\bar{\phi}_{B}|},\\
    &\mathcal{L}_{Y_{I_{B}} }(s) = 
     \mathrm{exp}\Big( 2\pi \lambda_B \int_{d_{00}}^{\sqrt{\frac{|\mathcal{A}|}{\pi}}} \big[
      \big(1+s\frac{T \rho_B \beta_{0} r^{-\alpha_1} }{\theta_{S_0}} \big)^{-|\bar{\phi}_{B}|}-1
      \big] r\mathrm{d}r   \Big).
  \end{aligned}$}
 \end{equation}

\end{lemma}

Based on   Lemma \ref{lemcoverage}, the analysis of the coverage probability is 
transformed into the analysis of the higher-order derivatives,
and the coverage of probability is rewritten as 
    \begin{equation}\label{eq:outtolap2}
      \begin{aligned}
        p_{\mathrm{c}}(d_{00})
     &=\sum_{i=0}^{k_{S_0}-1} \frac{(-1)^i}{i!} \frac{\partial^i}{\partial^i s}
         \Big\{ L(s )  \Big\}_{s=1},
      \end{aligned}
    \end{equation}
where
\begin{equation}
  \begin{aligned}
  &L(s ) = e^{-s\frac{TI_{e}}{\theta_{S_0}}} \cdot
  \Big(
    1+s\frac{T\rho_B\beta_0 d_{00}^{-\alpha_1}}{\theta_{S_0}}\Big)^{1-|\bar{\phi}_{B}|} 
    \\ &\cdot
    \mathrm{exp}\Big( 2\pi \lambda_B \int_{d_{00}}^{\sqrt{\frac{|\mathcal{A}|}{\pi}}} \big[
      \big(1+s\frac{T \rho_B \beta_{0} r^{-\alpha_1} }{\theta_{S_0}} \big)^{-|\bar{\phi}_{B}|}-1
      \big] r\mathrm{d}r   \Big).    
    \end{aligned}
\end{equation}

The higher-order derivatives of $L(s )$ is derived in the next part.

\subsection{Evaluation of  Higher-order Derivatives} 

The objective function $L(s )$ of the higher-order derivatives 
in   (\ref{eq:outtolap2}) is rewritten in the form of the exponential function, i.e.,
\begin{equation}
  \begin{aligned}
  L(s ) &= \mathrm{exp}\bigg\{\underbrace{-s\frac{TI_{e}}{\theta_{S_0}}}_{D_1(s)}+
  \underbrace{(1-|\bar{\phi}_{B}|) \mathrm{ln}\Big(
    1+s  T_{\theta_{S_0}} d_{00}^{-\alpha_1}  \Big)}_{D_2(s)} 
     \\ &+
     \underbrace{2\pi \lambda_B \int_{d_{00}}^{\sqrt{\frac{|\mathcal{A}|}{\pi}}} \big[
      \big(1+s T_{\theta_{S_0}}  r^{-\alpha_1}   \big)^{-|\bar{\phi}_{B}|}-1
      \big] r\mathrm{d}r }_{D_3(s)}
    \bigg\}    ,
  \end{aligned}
\end{equation}
where $T_{\theta_{S_0}}=\frac{T\rho_B\beta_0  }{\theta_{S_0}}$ is applied for convenience.

Since $ L(s )$ is a composite function of 
$g(s)=D_1(s)+D_2(s)+D_3(s)$, the special case of
Fa{\`a} di Bruno's formula with exponential functions can be applied to efficiently derive 
the $i$th order derivatives of $ L(s )$ \cite{Lyu2021IRS,johnson2002curious}, i.e.,
\begin{equation}
  \resizebox{1\hsize}{!}
  {$
  \begin{aligned}
    \frac{\partial^i}{\partial^i s}
         L(s ) &= \frac{\partial^i}{\partial^i s}\big\{\mathrm{exp}\big(
          g(s)\big)\big\} 
           = \mathrm{exp}\big(
            g(s)\big) B_i 
          \Big(\frac{\partial^1  g(s) }{\partial^1 s}, ...,\frac{\partial^i g(s) }{\partial^i s}\Big),
  \end{aligned}$}
\end{equation}
where $B_i(x_1,...,x_i)$ denotes the $i$th complete exponential Bell polynomial,
whose coefficients can be efficiently obtained 
according to its definition \cite{Tanbourgi2014dual,Ivanoff1958problem}.
The remaining work is to evaluate the higher-order derivatives of $g(s)$, which can be decomposed as
\begin{equation}
  \frac{\partial^i g(s) }{\partial^i s}=\frac{\partial^i D_1(s)}{\partial^i s}+\frac{\partial^i D_2(s)}{\partial^i s}+\frac{\partial^i D_3(s)}{\partial^i s}.
\end{equation}

For $D_1(s)$, the derivatives from order 1 to order $(k_{S_0}-1)$ can be expressed respectively as
\begin{equation}\label{eq:deltaD1}
  \frac{\partial^i D_1(s)}{\partial^i s}=\left \{ 
    \begin{aligned}
      -\frac{TI_{e}}{\theta_{S_0}}, \quad i=1\\
      0, \quad i>1
    \end{aligned}\right 
    .
\end{equation}

Additionally, for $D_2(s)$, there is 
\begin{equation}\label{eq:deltaD2}
  \frac{\partial^i D_2(s)}{\partial^i s}=(-1)^{i-1}(1-|\bar{\phi}_{B}|) (i-1)!
  \Big(\frac{T_{\theta_{S_0}} d_{00}^{-\alpha_1}}{1+T_{\theta_{S_0}} d_{00}^{-\alpha_1}s}\Big)^i.
\end{equation}

Finally, the   higher-order derivatives of $D_3(s)$ is 
\begin{equation}\label{eq:deltaD3}
  \begin{aligned}
  &\frac{\partial^i D_3(s)}{\partial^i s}=\\
  &2\pi \lambda_B \int_{d_{00}}^{\sqrt{\frac{|\mathcal{A}|}{\pi}}}  
      \frac{(|\bar{\phi}_{B}|+i-1)!}{(|\bar{\phi}_{B}|-1)!} \cdot
      \frac{(-T_{\theta_{S_0}} r^{-\alpha_1})^i}{(1+T_{\theta_{S_0}} r^{-\alpha_1}s)^{|\bar{\phi}_{B}|+i}}
        r\mathrm{d}r.
    \end{aligned}
\end{equation}

Finally, the network coverage probability can be obtained by substituting 
(\ref{eq:outtolap2}) with (\ref{eq:deltaD1}), (\ref{eq:deltaD2}) and (\ref{eq:deltaD3}) 
back into (\ref{eq:cov1}).

\section{Simulation Results}
In this section, the analytical results of the coverage probability of 
the hybrid cell and cell-free network are verified 
by the comparison with the Monte-Carlo (MC) simulation results.
Each result of the MS simulation is averaged from 1000 
randomly generated wireless node distributions
with 5   realizations per channel.
All the wireless nodes are randomly distributed in a circular area of radius 500m and UE 0 
is located at the center of the circle.
The densities of BSs, APs and UE are $\lambda_B=40/\mathrm{km}^2$,
$\lambda_A=200/\mathrm{km}^2$ and $\lambda_U=160/\mathrm{km}^2$,
respectively. Other relevant parameters are as follows:
$\alpha_1=2.7$, $\alpha_2=1.8$, $\frac{P_B}{\sigma^2}=130$dB, $P_A=3\times10^{-5} P_B$,
$N_B=4$, $N_A=2$, $C=3\times10^8$m/s, $f=3.5$GHz and $\beta_0=\delta_0=(\frac{C}{4\pi f})^2$.

\begin{figure}[!t]
  \centering
    {\includegraphics[width=0.83\columnwidth]{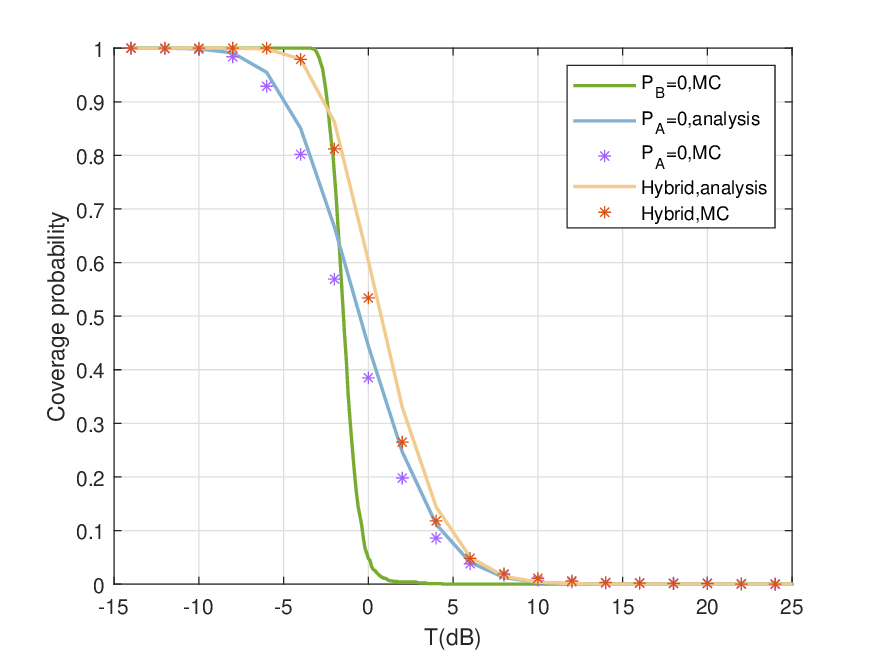}}%
  \caption{ Coverage probability of different architectures under different $T$.}
  \label{fig:Tpcov2}
\end{figure}

Fig. \ref{fig:Tpcov2} shows the coverage probability in   cellular networks ($P_A=0$), 
cell-free networks ($P_B=0$), and hybrid cellular and cell-free networks for different SINR thresholds $T$.
The  coverage probability  
is obtained from the linear weighted probability of the 
upper and lower integers of $k_{\theta_S}$, i.e.,
\begin{equation}
  p_{\mathrm{c }} =(\left\lceil k_{\theta_S}\right\rceil -
  k_{\theta_S})p_{\mathrm{c ,\left\lfloor k_{\theta_S} \right\rfloor }} 
+( 
k_{\theta_S}-\left\lfloor k_{\theta_S} \right\rfloor )p_{\mathrm{c ,
\left\lceil k_{\theta_S}\right\rceil } }.
\end{equation}
From Fig. \ref{fig:Tpcov2}, It is expected that  the coverage probability analysis of the hybrid network 
is generally consistent with the results of MC simulation under different SINR threshold, 
and can also be applied to the special case where $P_A=0$.
Compared with traditional cellular networks, hybrid networks effectively improve 
the communication performance of edge UE and reduce the performance gaps 
between UE. Compared with cell-free networks, such hybrid networks can 
achieve higher peak SINR.
Therefore, by deploying low-power APs, hybrid cellular and cell-free networks can provide 
UE with uniformly good communication services while obtaining better peak SINR performance.

\section{Conclusion}
In this paper,  
the hybrid cellular and cell-free 
network is modeled by the stochastic geometry approach,
revealing the coupling of the signal and interference from both
the cellular and cell-free networks.
Moment matching is used to approximate the aggregate signal received from 
the hybrid network to address the difficulty of distribution analysis 
due to conjugate beamforming.
The coverage probability  is then obtained 
by the Laplace transform for interference.
The analysis of the coverage probability of  hybrid networks
is validated by MC simulation, demonstrating that hybrid networks 
can reduce the  performance gap while improving the peak SINR performance.
 
\section{Acknowledgment}
This work was supported by the National Key R\&D Program
of China with Grant number 2019YFB1803400,
and  by the National Natural Science Foundation of China with grant number 62071114.

\bibliographystyle{IEEEtran}

\bibliography{IEEEabrv,ref}


\end{document}